\documentclass[pdflatex,sn-nature]{sn-jnl}

\usepackage{float}
\usepackage{graphicx}
\usepackage{amsmath,amssymb,bm,mathtools}
\usepackage{booktabs,array,multirow}
\usepackage{xcolor}
\usepackage{hyperref}
\usepackage{placeins}

\newcommand{\PopulationSize}{500}
\newcommand{\AssessmentErrorMin}{0}
\newcommand{\AssessmentErrorMax}{0.10}
\newcommand{\AssessmentErrorStep}{0.01}
\newcommand{\InitialBadFraction}{0.5}
\newcommand{\FinalMCTime}{998}

\newcommand{\NumberRuns}{50}

\newcommand{\safeincludegraphics}[2][]{%
  \IfFileExists{fig/#2}{\includegraphics[#1]{fig/#2}}{%
    \fbox{\parbox[c][4.3cm][c]{0.92\linewidth}{%
      \centering\textbf{Figure placeholder}\\[0.5em]
      Upload file: \texttt{\detokenize{fig/#2}}%
    }}%
  }%
}

\newcommand{\Good}{\mathrm{G}}
\newcommand{\Bad}{\mathrm{B}}
\newcommand{\Coop}{\mathrm{C}}
\newcommand{\Defect}{\mathrm{D}}

\newcommand{\Pp}{\mathbb{P}}

\begin{document}

\title[Pairwise correlations in private assessment]{Pairwise opinion correlations improve predictions for the Leading Eight under private assessment}
\title[Pairwise correlations in private assessment]{Opinion correlations shape cooperation under private assessment in indirect reciprocity}
\title[Pairwise correlations in private assessment]{Pairwise assessment correlations predict cooperation of the Leading Eight under private assessment}

\author[1]{\fnm{Jong-Won} \sur{Hong}}
\author[1,2]{\fnm{Sunhee} \sur{Chae}}
\author*[1,2]{\fnm{Hye Jin} \sur{Park}}\email{hyejin.park@inha.ac.kr}

\affil[1]{\orgdiv{Department of Physics}, \orgname{Inha University},
\orgaddress{\city{Incheon}, \postcode{22212}, \country{Republic of Korea}}}
\affil[2]{\orgdiv{Institute of Quantum Science}, \orgname{Inha University},
\orgaddress{\city{Incheon}, \postcode{22212}, \country{Republic of Korea}}}

\abstract{
Indirect reciprocity can sustain cooperation when individuals use reputations to decide whom to help. The Leading Eight norms sustain such cooperation under public assessment, where all individuals share the same assessments. Under private assessment, however, observers may disagree, yet their assessments remain correlated. Two quantities, the average fraction of good assessments $h$ and the conditional probability $h_{\mathrm G}$ that one observer assesses a player as good given that another does, are known to determine the stability of cooperation, but how $h_{\mathrm G}$ itself is determined has remained open. Within a pairwise approximation, we derive a closed equation for $h_{\mathrm G}$ that, together with that for $h$, determines both quantities for all eight norms, including third-order ones. This description substantially improves agreement with simulations over a mean-field approximation that neglects correlations, and recovers the known classification of the Leading Eight into three classes. The remaining discrepancies are largest for L2 and L5 at low error rates and decrease as assessment error increases. Relating $h_{\mathrm G}$ to the variance of individual goodness suggests that these discrepancies reflect information beyond the first two moments of the goodness distribution. Our framework provides a concise and analytically tractable tool for evaluating social norms under private assessment.
}


\maketitle
\raggedbottom
\section{Introduction}
\label{sec:introduction}
Cooperation requires an individual to incur a cost to benefit another, whereas defection does not. Defectors therefore gain an immediate advantage in a one-shot interaction, yet cooperation is commonly observed. To explain this puzzle, several evolutionary mechanisms have been proposed, including kin selection, direct reciprocity, indirect reciprocity, network reciprocity, and group selection~\cite{Nowak2006Science}. Among these, indirect reciprocity is particularly relevant in large populations, where interactions frequently occur between strangers. There, helping another individual can improve the helper's reputation and thereby increase the likelihood of receiving help from others~\cite{NowakSigmund1998,NowakSigmund2005,WedekindMilinski2000,MilinskiEtAl2001}. In a typical indirect-reciprocity interaction, one individual, the donor, decides whether to help another individual, the recipient, while third parties may observe the interaction and update the donor's reputation.

In such interactions, social norms are typically described in terms of an assessment rule and an action rule~\cite{Sigmund2012,BrandtSigmund2004}. The assessment rule, which constitutes the core of a norm, prescribes how observers update the donor's reputation following the interaction, for instance whether defection against a bad recipient is judged good or bad. The action rule then determines whether a donor cooperates with a recipient, given their reputations. These norms take reputations to be simply good or bad, and observers judge the donor from its action together with the reputations of both players. Within this class, an exhaustive search identified eight norms, collectively known as the Leading Eight, that sustain cooperation and resist invasion by behavioral mutants under public assessment, where every observer holds the same assessment of any given individual~\cite{OhtsukiIwasa2004,OhtsukiIwasa2006}. 

The Leading Eight largely agree on the action rule: a donor always cooperates with a good recipient, and a good donor defects against a bad one. The only exception is a bad donor facing a bad recipient. Beyond this, they are distinguished by their assessment rules, which nonetheless share a common logic: cooperation with a good recipient is judged good, unjustified defection against a good recipient is judged bad, defection against a bad recipient by a good donor is justified, and individuals with damaged reputations can regain good standing by cooperating with good recipients~\cite{PanchanathanBoyd2003,PanchanathanBoyd2004,OhtsukiIwasa2006}. Together, these features allow cooperation to persist even when individuals occasionally defect or make mistakes, since a damaged reputation can be repaired rather than punished forever. This robustness is why the Leading Eight have become a standard benchmark in the study of indirect reciprocity. Subsequent work has extended the framework to conditional moral assessment, more complex dependence on past reputations, local reputation and selection, multilevel reputation states, and stochastic or dual reputation updating~\cite{SasakiOkadaNakai2017,SantosEtAl2018,PodderRighiTakacs2021,MuraseKimBaek2022,MuraseHilbe2023}.

This success, however, rests on an idealization. Public assessment assumes that everyone shares the same assessment of each individual. Real populations rarely meet this condition. Instead, each observer forms its own assessment of every other individual, a setting known as private assessment~\cite{Okada2018}. Independent assessment errors then lead observers to disagree about the same player~\cite{Uchida2010,UchidaSasaki2013,Hilbe2018,Schmid2021}, while observing the same actions under a shared assessment rule induces statistical correlations among their assessments~\cite{MuraseHilbe2024}. Under private assessment, these dynamics can reshape both the structure of assessments in the population and the evolutionary stability of social norms~\cite{FujimotoOhtsuki2023,Schmid2023,OkadaDeSilva2024,MuraseHilbe2024}.

Indeed, several norms of the Leading Eight have been shown to fail to sustain cooperation under private assessment~\cite{Hilbe2018,FujimotoOhtsuki2024}. Two recent studies analyze this problem through an individual's goodness, defined as the fraction of observers who assess that individual as good. Fujimoto and Ohtsuki characterized the stationary distribution of goodness and showed that the Leading Eight fall into three classes with distinct assessment structures and levels of cooperation~\cite{FujimotoOhtsuki2024}. Although this approach captures the full distribution, the distribution is determined only implicitly by an integral equation, which can be solved analytically for only a few norms and otherwise requires numerical iteration. Murase and Hilbe instead determined evolutionary stability from two quantities: the population-averaged goodness $h$ and the conditional probability $h_{\mathrm G}$ that an observer assesses a player as good, given that another observer assesses the same player as good, both of which follow from the first two moments of the goodness distribution~\cite{MuraseHilbe2024}. Their steady-state condition, however, provides only a single relation between $h$ and $h_{\mathrm G}$, yielding $h$ as a function of $h_{\mathrm G}$; how $h_{\mathrm G}$ itself is determined in a given social setting was left as an open question~\cite{MuraseHilbe2024}.

Here we address this question for the Leading Eight under private assessment. Within a pairwise approximation, we derive a closed equation for $h_{\mathrm G}$ that, together with the equation for $h$, determines both quantities without input from simulations. The resulting predictions are substantially more accurate than a mean-field description that neglects correlations among assessments. Unlike previous approaches, which either are restricted to second-order norms~\cite{MuraseHilbe2024} or require numerical solution of an integral equation for most of the Leading Eight~\cite{FujimotoOhtsuki2024}, our framework applies to all eight norms, including third-order ones. It recovers the three classes identified by Fujimoto and Ohtsuki, (L1, L3, L4, L7), (L2, L5), and (L6, L8), together with the central role of how a norm judges cooperation with a bad recipient~\cite{FujimotoOhtsuki2024}. By reducing the assessment structure to two quantities, our approach offers a concise and analytically tractable tool for evaluating social norms under private assessment.

\section{Model}
\label{sec:model}
\subsection{Image matrix and its update rule}
\label{sec:private-image-matrix}

\begin{figure}[!htbp]
\centering
\safeincludegraphics[width=0.94\linewidth]{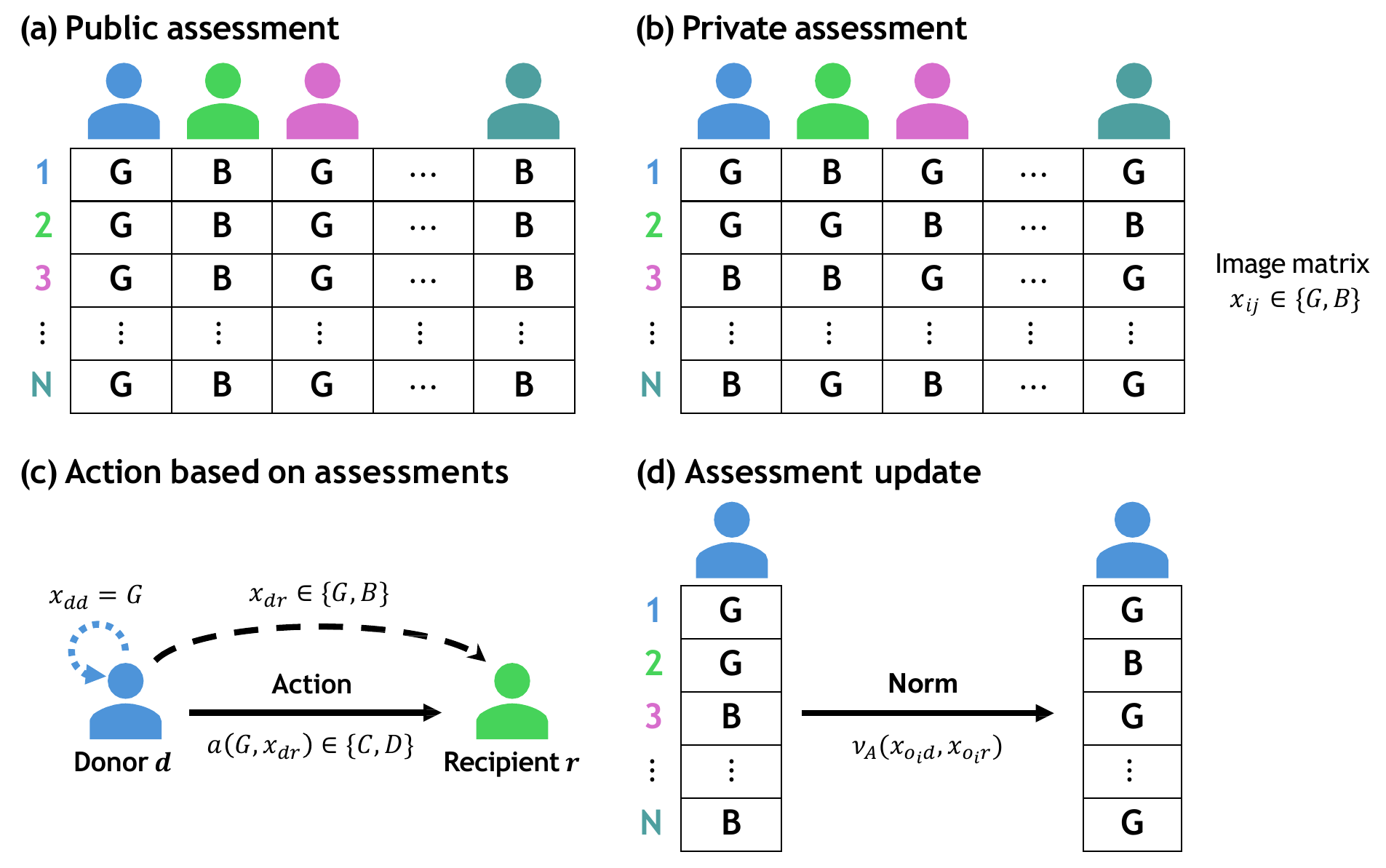}
\caption{\textbf{Image matrix, action rule, and assessment update.}
(\textbf{a}, \textbf{b}) The image matrix $\bm{x}$, whose element $x_{ij}$ represents player $i$'s assessment of player $j$. (\textbf{a}) Under public assessment, all players share a common assessment of each player. (\textbf{b}) Under private assessment, each player holds its own assessment of every other player. (\textbf{c}) In each interaction, a donor $d$ cooperates with or defects against a recipient $r$ according to its own assessment of the recipient, with the donor's self-assessment fixed to $\Good$. (\textbf{d}) Each observer then updates its assessment of the donor according to the assessment rule in Table~\ref{tab:leading-eight}, which depends on the donor's action and on the observer's pre-update assessments of the donor and the recipient. With probability $\epsilon$, each observer independently assigns the opposite of the prescribed assessment.
}
\label{fig:model}
\end{figure}

To examine how assessments are formed under private assessment, we consider the donation game, in which a donor decides whether to cooperate with a recipient, and the other individuals, including the recipient, observe the interaction and update their assessments of the donor. We consider a well-mixed population of $N$ individuals, in which every pair can interact as donor and recipient. The binary variable $x_{ij}\in\{0,1\}$ denotes observer $i$'s assessment of player $j$, with $1$ denoting good ($\Good$) and $0$ denoting bad ($\Bad$). The assessments of the whole population are represented by the image matrix $\bm{x}$ with elements $x_{ij}$; its rows correspond to observers and its columns to the individuals being assessed, see Fig.~\ref{fig:model}(a)-(b). We assume that all individuals consider themselves $\Good$ throughout the dynamics, so that $x_{ii}=1$ for all individuals at all times. We have numerically confirmed that our results are robust when this constraint is relaxed (see SI). 

In each interaction, a donor--recipient pair $(d,r)$ with $d\neq r$ is chosen at random. The donor chooses its action $A$ according to the action rule $a(x_{dd},x_{dr})$ specified in Table~\ref{tab:leading-eight}, where $A=1$ denotes cooperation and $A=0$ denotes defection, also see Fig.~\ref{fig:model}(c). Since self-assessments are fixed to $1$, the action rule reduces to $A=a(1,x_{dr})$: the donor cooperates if and only if it assesses the recipient as $\Good$. Throughout, we assume that actions are free of errors.

\begin{table}[!h]
\caption{\textbf{The Leading Eight assessment and action rules.}
Entries in the first two blocks give the assessment assigned to the donor after cooperation or defection, respectively, as a function of the observer's pre-update assessment of the donor $x_{od}$ and the observer's assessment of the recipient $x_{or}$. Thus, within the assessment-rule blocks, the first and second symbols in each column correspond to $x_{od}$ and $x_{or}$, respectively. Entries in the final block give the donor's intended action as a function of the donor's self-assessment $x_{dd}$ and the donor's assessment of the recipient $x_{dr}$. Accordingly, the $\Good\Bad$ column under $a(x_{dd},x_{dr})$, for example, represents $a(\Good,\Bad)$. C and D in the action block denote cooperation and defection, respectively.
}
\label{tab:leading-eight}
\centering
\scriptsize
\setlength{\tabcolsep}{2.4pt}
\begin{tabular}{@{}c cccc cccc cccc@{}}
\toprule
& \multicolumn{4}{c}{$\nu_{\Coop}(x_{od},x_{or})$}
& \multicolumn{4}{c}{$\nu_{\Defect}(x_{od},x_{or})$}
& \multicolumn{4}{c}{$a(x_{dd},x_{dr})$}\\
\cmidrule(lr){2-5}
\cmidrule(lr){6-9}
\cmidrule(lr){10-13}
Norm
& $\Good\Good$ & $\Good\Bad$ & $\Bad\Good$ & $\Bad\Bad$ 
& $\Good\Good$ & $\Good\Bad$ & $\Bad\Good$ & $\Bad\Bad$
& $\Good\Good$ & $\Good\Bad$ & $\Bad\Good$ & $\Bad\Bad$\\
\midrule
L1 & G&G&G&G & B&G&B&B & C&D&C&C\\
L2 & G&B&G&G & B&G&B&B & C&D&C&C\\
L3 & G&G&G&G & B&G&B&G & C&D&C&D\\
L4 & G&G&G&B & B&G&B&G & C&D&C&D\\
L5 & G&B&G&G & B&G&B&G & C&D&C&D\\
L6 & G&B&G&B & B&G&B&G & C&D&C&D\\
L7 & G&G&G&B & B&G&B&B & C&D&C&D\\
L8 & G&B&G&B & B&G&B&B & C&D&C&D\\
\bottomrule
\end{tabular}
\end{table}

Following the interaction, every observer $o\neq d$ updates its assessment of the donor according to the assessment rule $\nu_A(x_{od},x_{or})\in\{0,1\}$. Note the difference in perspective between the two rules: the action rule uses the donor's own assessments, because the donor chooses the action, whereas the assessment rule uses each observer's assessments, because each observer forms its own judgment. Consequently, the same action can be judged differently by observers who disagree about the donor or the recipient. The form of the assessment rule $\nu_A(x_{od},x_{or})$ covers the Leading Eight in general: the prescribed assessment may depend on the donor's action $A$ together with the observer's pre-update assessments of the donor, $x_{od}$, and of the recipient, $x_{or}$. Norms that do not use $x_{od}$ are second-order (L3 and L6), and the remaining ones are third-order. Each observer applies the rule independently and, with probability $\epsilon$, assigns the opposite assessment to the prescribed one. Hence, for pre-update inputs $u=x_{od}$ and $v=x_{or}$, the probability that the donor is assessed as $\Good$ after action $A$ is
\begin{equation}
 \mu_A(u,v) = \epsilon+(1-2\epsilon)\nu_A(u,v).
\end{equation}
The entries of the donor's column in the image matrix are updated simultaneously, while the self-assessment $x_{dd}$ remains fixed at $\Good$ and is not subject to assessment errors, see Fig.~\ref{fig:model}(d). We assume $0\leq\epsilon\leq1/2$ throughout.

\subsection{Goodness and assessment correlation }

To characterize how each player is assessed by others, we use the goodness of player $j$, $p_j=(N-1)^{-1}\sum_{i(\neq j)}x_{ij}$, defined as the fraction of observers who assess $j$ as $\Good$. The population-averaged goodness is then
\begin{equation}
 h=\frac{1}{N}\sum_{j=1}^{N}p_j=\frac{1}{N(N-1)}\sum_{i\neq j}x_{ij}.
 \label{eq:h-def}
\end{equation}
Here and below, $\sum_{i(\neq j)}$ denotes a sum over $i$ for fixed $j$, whereas $\sum_{i\neq j}$ denotes a sum over all ordered pairs $(i,j)$ with $i\neq j$. Because each player always assesses itself as $\Good$, the action rule reduces to cooperating with recipients assessed as $\Good$ and defecting against those assessed as $\Bad$. Hence, $h$ also equals the average probability that a donor cooperates. If different observers' assessments of the same player are assumed to be statistically independent, the dynamics of $h$ closes in terms of $h$ alone. We refer to this closure as the one-variable mean-field (MF) approximation.

This independence assumption, however, generally fails. Observers update their views of player $j$ based on the same actions of $j$, so a player assessed as $\Good$ by one observer is more likely to be assessed as $\Good$ by another. To capture this correlation, we also use the population-averaged conditional probability $h_{\mathrm G}$ that a player is assessed as $\Good$ by one observer, given that another observer assesses the same player as $\Good$:
\begin{equation}
 h_{\mathrm G}
 =   \frac{\sum_{j}p_j\left((N-1)p_j-1\right)}{\sum_{j}(N-2)p_j}.
\label{eq:hG-matrix}
\end{equation}
This conditional-probability description follows the opinion-synchronization framework of Murase and Hilbe~\cite{MuraseHilbe2024}.

For any three distinct players $i$, $j$, and $k$, we then approximate
\begin{equation}
 \Pp(x_{ij}=1)\approx h, \qquad
 \Pp(x_{kj}=1\mid x_{ij}=1)\approx h_{\mathrm G}.
 \label{eq:h-hG-def}
\end{equation}
That is, we replace these probabilities by their population averages. Thus, within the MF approximations in Eq.~\eqref{eq:h-hG-def}, relabelling of individuals is exchangeable. We derive the coupled dynamics of $h$ and $h_{\mathrm G}$ and obtain their stationary values. To validate the one- and two-variable MF approximations, we compare their predictions for the stationary $h$ with Monte Carlo simulations, in which $h$ is computed directly from the image matrix. The simulation protocol and the analytical derivations are detailed in Section~\ref{sec:methods}.

\section{Results}
\label{sec:results}
\begin{figure}[!htbp]
\centering
\safeincludegraphics[width=0.96\linewidth]{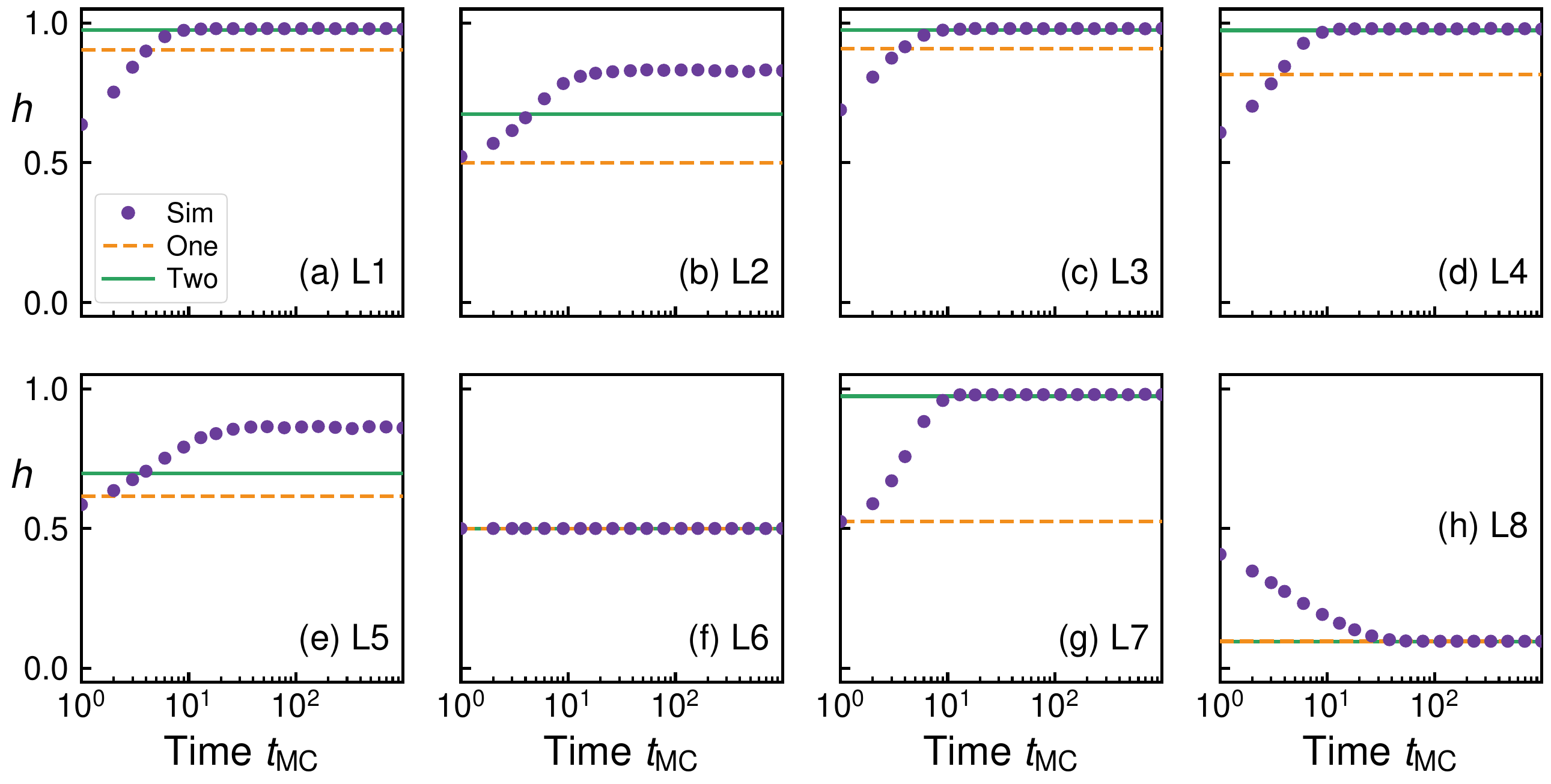}
\caption{\textbf{Time evolution of the population-averaged goodness and theoretical predictions of its stationary value.}
The population-averaged goodness $h$ is shown as a function of Monte Carlo time $t_{\mathrm{MC}}$ (logarithmic scale) for L1--L8 at assessment error $\epsilon=0.01$. Purple circles (``Sim'') show simulation results averaged over $50$ independent realizations with population size $N=500$, starting from an initial fraction $0.5$ of $\Bad$ assessments. Orange dashed lines (``One'') and green solid lines (``Two'') indicate the stationary values $\tilde h$ predicted by the one- and two-variable MF approximations, respectively. The two-variable approximation is closer to the simulations than the one-variable approximation for all norms and is in excellent agreement except for L2 and L5.}
\label{fig:time_h}
\end{figure}

The Leading Eight were originally identified by their ability to sustain cooperation under public assessment. Under private assessment, however, observers can disagree in their assessments, and cooperation may not persist. Since a donor cooperates only with a recipient it regards as $\Good$, the population-averaged goodness $h$ directly gives the level of cooperation in our model, and its time evolution reveals how robust each norm is under private assessment. As shown in Fig.~\ref{fig:time_h}, the trajectories of $h$ differ across norms and converge to different stationary levels, consistent with previous work~\cite{FujimotoOhtsuki2024}. To analytically predict the stationary value of $h$, denoted by $\tilde h$, we apply a mean-field (MF) approximation. In particular, within the two-variable MF approximation, we derive the stationary condition for $h_{\mathrm G}$ and close it by approximating higher-order correlations in terms of $h$ and $h_{\mathrm G}$.

In the stationary state, $h$ is unchanged by assessment updates. After each interaction, every observer updates its assessment of the donor, and the probability $F$ that an updated assessment is $\Good$ must therefore equal the goodness itself, $\tilde h=F$. This probability is determined by three assessments: the observer's assessments of the donor, $x_{od}$, and of the recipient, $x_{or}$, and the donor's assessment of the recipient, $x_{dr}$, which determines the donor's action.

In the one-variable MF approximation, the three assessments are treated as independent, each being $\Good$ with probability $h$. Then $F$ reduces to a function of $h$ alone,
\begin{equation}
 F(h)=
 \sum_{x_{od}, x_{or},x_{dr}\in\{0,1\}}
 \pi_h(x_{od})\,\pi_h(x_{or})\,\pi_h(x_{dr})\, \mu_{A}(x_{od}, x_{or}),
 \label{eq:MF-h-closed}
\end{equation}
where $\pi_h(x)=h^x(1-h)^{1-x}$ is the probability that a player assesses another player as $x\in\{0,1\}$ (see Methods in Sec.~\ref{sec:methods} for the derivation). The stationary value $\tilde h$ in the one-variable MF approximation is obtained by solving the self-consistency equation $\tilde h=F(\tilde h)$.

However, the independence of $x_{or}$ and $x_{dr}$ does not hold in general. Assessments of the same player by different observers are correlated, because they are updated upon observing the same action. 
To take this correlation into account, we use the conditional probability $h_{\mathrm G}$ ($h_{\mathrm B}$) that an observer regards a player as $\Good$ given that another observer regards the same player as $\Good$ ($\Bad$), and write the joint probability as $\Pp(x_{or}, x_{dr}) = \pi_h(x_{dr})\,\pi(x_{or}\,|\,x_{dr})$, where $\pi(x\,|\,1)=\pi_{h_{\mathrm G}}(x)$ and $\pi(x\,|\,0)=\pi_{h_{\mathrm B}}(x)$. We then obtain the two-variable MF expression
\begin{equation}
 F(h, h_{\mathrm G})=
 \sum_{x_{od}, x_{or},x_{dr}\in\{0,1\}}
 \pi_h(x_{od})\,\pi_h(x_{dr})\,\pi(x_{or}\,|\,x_{dr})\, \mu_{A}(x_{od}, x_{or}).
 \label{eq:MF-g-two}
\end{equation}
Since $h_{\mathrm B}=h(1-h_{\mathrm G})/(1-h)$, $F$ depends only on $h$ and $h_{\mathrm G}$. To determine $\tilde h$ from $\tilde h =F(\tilde h, \tilde h_{\mathrm G})$, we thus need a second condition for $\tilde h_{\mathrm G}$.

This condition follows from the joint probability $h\,h_{\mathrm G}$ that two distinct observers, labeled 1 and 2, both regard the same player as $\Good$. In the stationary state, it must equal the probability $Q$ that both observers regard the donor as $\Good$ after an update. Evaluating $Q$ requires the joint probability of the three assessments of the recipient, $x_{dr}$, $x_{1r}$, and $x_{2r}$. To close the equations, we approximate
\begin{equation}
 \Pp(x_{dr},x_{1r}, x_{2r})\approx \Pp(x_{dr})\,\Pp(x_{1r}\,|\,x_{dr})\,\Pp(x_{2r}\,|\,x_{dr}),
 \label{eq:three-point}
\end{equation}
that is, we assume $\Pp(x_{2r}\,|\,x_{1r},x_{dr})\approx \Pp(x_{2r}\,|\,x_{dr})$. Then $Q$ is also a function of $h$ and $h_{\mathrm G}$,
\begin{equation}
\begin{aligned}
 Q(h, h_{\mathrm G})=
 \sum_{\substack{x_{dr},x_{1d}, x_{1r},\\x_{2d}, x_{2r}\in\{0,1\}}}
 &\pi_h(x_{1d})\,\pi(x_{2d}\,|\,x_{1d})\,
 \pi_h(x_{dr})\,\pi(x_{1r}\,|\,x_{dr})\,\pi(x_{2r}\,|\,x_{dr})\\
 &\times \mu_{A}(x_{1d}, x_{1r})\,\mu_{A}(x_{2d}, x_{2r})
 \end{aligned}
 \label{eq:MF-hG-closed}
\end{equation}
(see Methods in Sec.~\ref{sec:methods}), and the stationary values are obtained by solving
\begin{equation}
 \tilde h = F(\tilde h, \tilde h_{\mathrm G}),
 \qquad
 \tilde h\, \tilde h_{\mathrm G} = Q(\tilde h, \tilde h_{\mathrm G}).
 \label{eq:stationary}
\end{equation}

Figure~\ref{fig:time_h} compares the stationary values predicted by the one- and two-variable MF approximations with the simulation results. Although crude, the one-variable approximation captures the results for L1 and L3 reasonably well at $\epsilon=0.01$, and in particular those for L6 and L8, which display low goodness, suggesting that assessments are nearly uncorrelated under these two norms. The two-variable approximation substantially improves the prediction, showing excellent agreement with the simulations except for L2 and L5. This shows that our closure for $h_{\mathrm G}$ performs well and that two-point correlations are sufficient to predict $\tilde h$ for those norms.

\begin{figure}[!htbp]
\centering
\safeincludegraphics[width=0.96\linewidth]{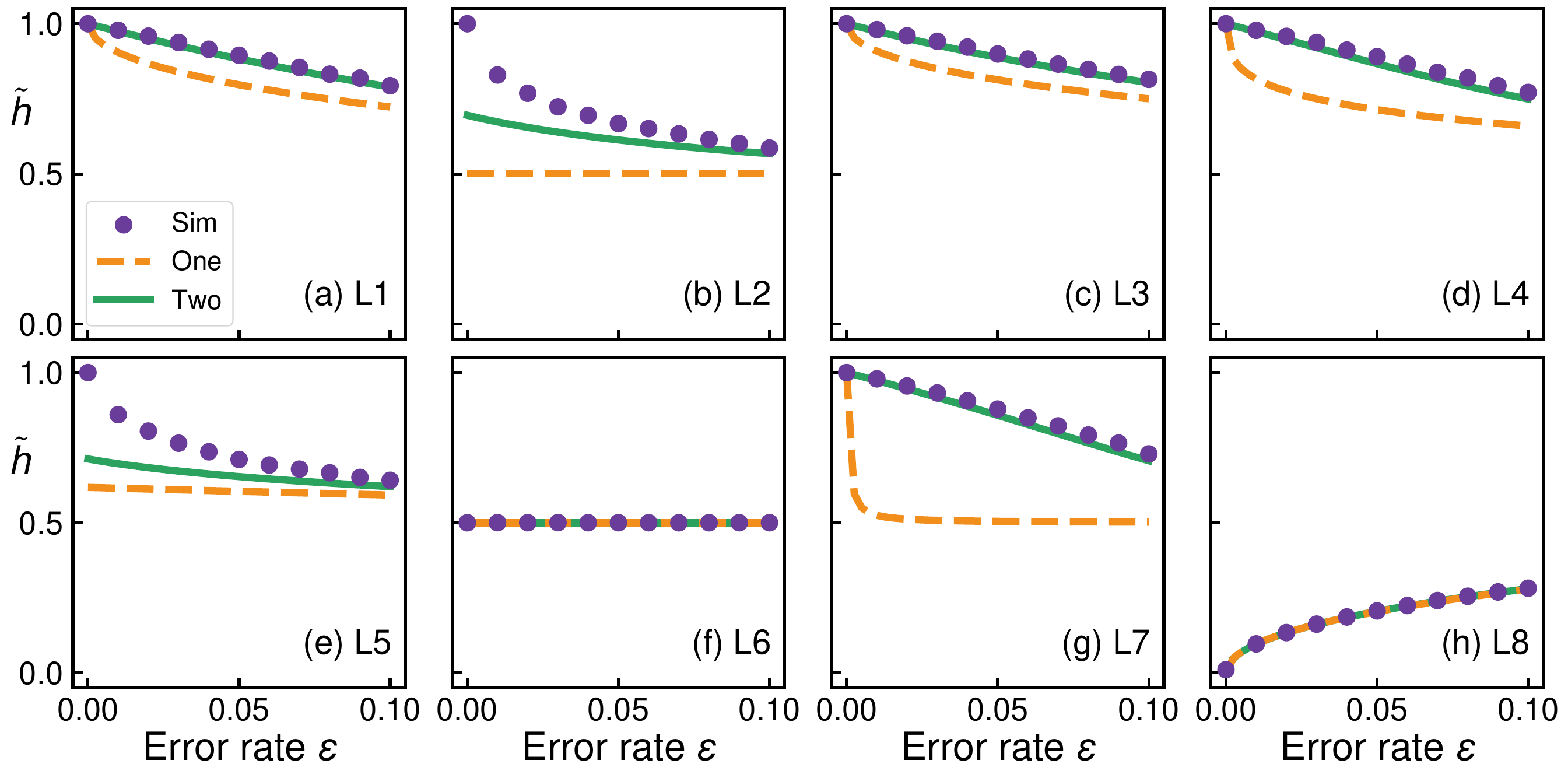}
\caption{\textbf{Stationary goodness $\tilde h$ as a function of the assessment error $\epsilon$.}
Purple circles (``Sim'') show simulation results for L1--L8, obtained as the ensemble average of $h$ at the final time over $50$ independent runs with population size $N=500$. Orange dashed curves (``One'') and green solid curves (``Two'') show the predictions of the one- and two-variable MF approximations, respectively. The one-variable approximation captures the overall trend of $\tilde h$ with $\epsilon$ but deviates systematically from the simulations, whereas the two-variable approximation agrees closely with them. Visible discrepancies remain only for L2 and L5.}
\label{fig:steady_h}
\end{figure}

To examine how the MF approximations perform for different assessment errors, we compare the predicted $\tilde h$ with the simulations while varying $\epsilon$. As shown in Fig.~\ref{fig:steady_h}, the one-variable approximation captures the trend of $\tilde h$ with $\epsilon$ but deviates systematically from the simulations. Incorporating the additional variable $h_{\mathrm G}$ substantially improves the agreement for L1, L3, L4, and L7, and also reduces the discrepancies for L2 and L5. These results indicate that two-point correlations are sufficient to predict $\tilde h$ for most norms, whereas the remaining discrepancies for L2 and L5, particularly at low assessment error, suggest that higher-order correlations become relevant there.

\begin{figure}[!htbp]
\centering
\safeincludegraphics[width=0.96\linewidth]{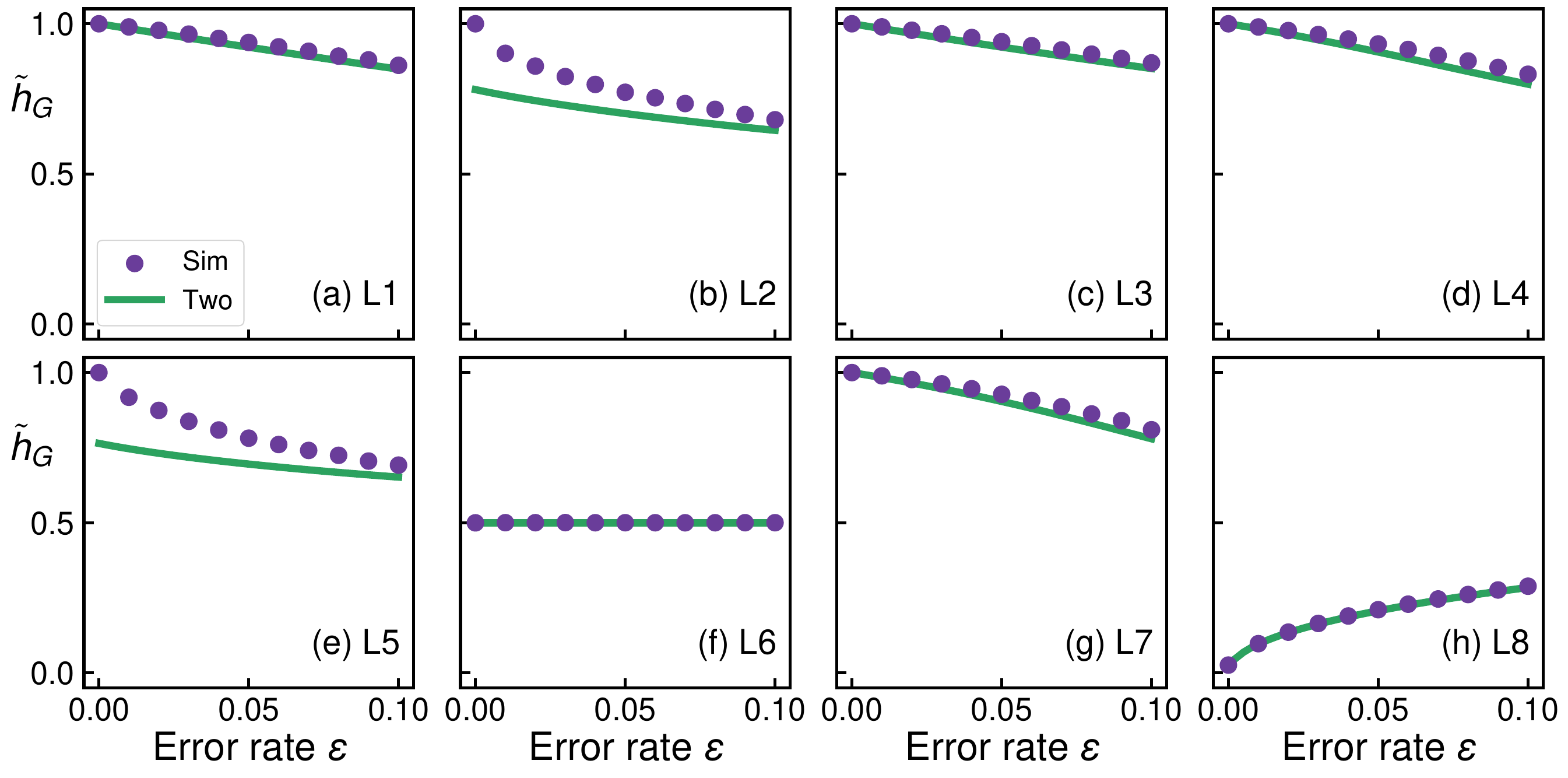}
\caption{\textbf{Stationary conditional probability $\tilde h_{\mathrm G}$ as a function of the assessment error $\epsilon$.}
Purple circles (``Sim'') show simulation results for L1--L8, obtained as the ensemble average of $h_{\mathrm G}$ at the final time over $50$ independent runs with population size $N=500$. Green solid curves (``Two'') show the predictions of the two-variable MF approximation, which reproduces both the magnitude of $\tilde h_{\mathrm G}$ and its dependence on $\epsilon$ for most norms. The largest deviations occur for L2 and L5 and decrease as $\epsilon$ increases.}
\label{fig:hG-comparison}
\end{figure}

The two-variable approximation also predicts $\tilde h_{\mathrm G}$, which is absent from the one-variable description. Figure~\ref{fig:hG-comparison} shows that the predicted $\tilde h_{\mathrm G}$ agrees closely with the simulations across the norms and assessment errors in most cases. The largest deviations again occur for L2 and L5, particularly at low assessment error, and decrease as the error increases. 

Comparison with Fig.~\ref{fig:steady_h} shows that $\tilde h_{\mathrm G}$ is generally larger than $\tilde h$ in the simulations, indicating a positive correlation between different observers' assessments of the same player. In addition, the difference between $h_{\mathrm G}$ and $h$ reflects how strongly observers agree in their assessments of the same player. We quantify this agreement by the covariance of two distinct observers' assessments of the same player $j$,
\begin{equation}
 C=\mathbb{E}(x_{1j}x_{2j})-\mathbb{E}(x_{1j})\,\mathbb{E}(x_{2j})=h\,h_{\mathrm G}-h^2=h(h_{\mathrm G}-h),
 \label{eq:cov}
\end{equation}
which vanishes when the observers assess independently, $h_{\mathrm G}=h$. For large $N$, the covariance equals the variance of the goodness $p_j$ across players~\cite{MuraseHilbe2024}, and thus measures how heterogeneously players are assessed (see SI).

\begin{figure}[!htbp]
\centering
\safeincludegraphics[width=0.96\linewidth]{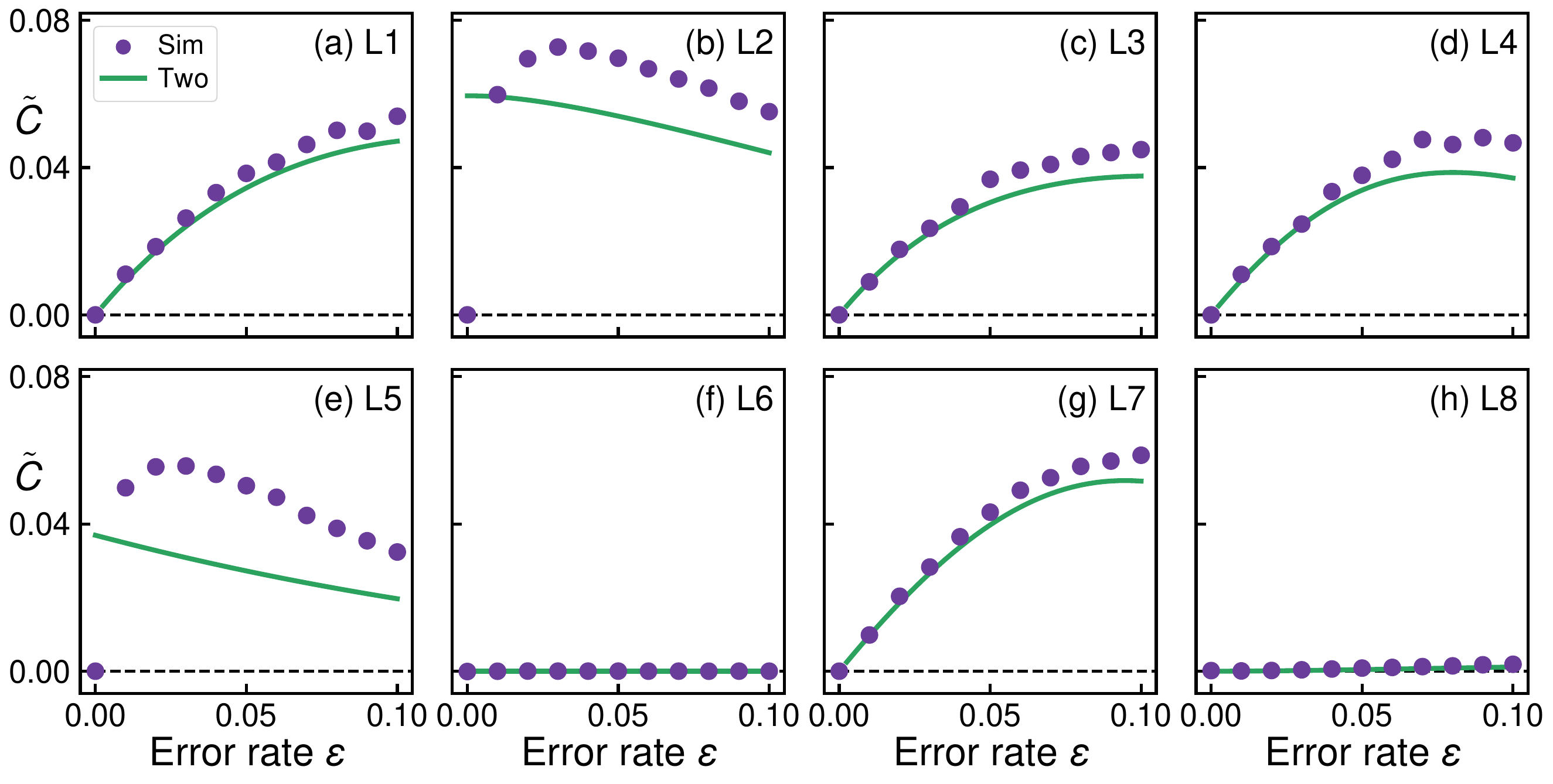}
\caption{\textbf{Covariance between observers' assessments of the same player as a function of the assessment error $\epsilon$. }
The stationary covariance $\tilde C$ between two distinct observers' assessments of the same player is shown for L1--L8. Purple circles (``Sim'') show simulation results, and green solid curves (``Two'') show the predictions of the two-variable MF approximation. The covariance nearly vanishes for L6 and L8, and the two-variable approximation deviates from the simulations mainly for L2 and L5.}
\label{fig:cov}
\end{figure}

Figure~\ref{fig:cov} shows the stationary covariance $\tilde C$ as a function of $\epsilon$. In the simulations, $\tilde C\geq0$ for all norms, indicating that observers' assessments of the same player are positively correlated. The covariance is nearly zero for L6 and L8, for which the one-variable approximation is already accurate, and is larger for the norms where the one-variable approximation fails. The two-variable approximation reproduces $\tilde C$ well for all norms except L2 and L5, particularly at low assessment error, which is consistent with the discrepancies in $\tilde h$ for these norms. This marks the limit of our approach, which we discuss below.

\section{Discussion}
\label{sec:discussion}
Under private assessment, predicting how much cooperation a social norm sustains requires accounting for correlations among observers' assessments. Murase and Hilbe showed that the average goodness $h$ and the conditional probability $h_{\mathrm G}$ determine the stability of cooperation, but left open how $h_{\mathrm G}$ itself is determined~\cite{MuraseHilbe2024}. We addressed this question for the Leading Eight by deriving the stationary condition for $h_{\mathrm G}$ and closing it with an approximation for the three-point probability as shown in Eq.~\eqref{eq:three-point}. In particular, our equations require no input from simulations and apply to all eight norms. Comparison with simulations shows that the one-variable MF approximation deviates systematically from them, whereas the two-variable approximation agrees closely with them except for L2 and L5, particularly at low assessment error.

As the variance of the goodness across players for large $N$ is $h(h_{\mathrm G}-h)$, which equals the covariance $C$ between two observers' assessments, tracking $h_{\mathrm G}$ in addition to $h$ gives the second moment of the goodness distribution. From this perspective, the Leading Eight fall into three groups. For L6 and L8, the variance nearly vanishes, and the one-variable approximation suffices. For L1, L3, L4, and L7, the variance is substantial but captured at the pairwise level. For L2 and L5, the pairwise description is insufficient. These groups coincide with the three classes identified from the full goodness distribution~\cite{FujimotoOhtsuki2024}, which are distinguished by how cooperation with a $\Bad$ recipient is assessed (Table~\ref{tab:leading-eight}). Our two equations recover this classification without solving for the distribution itself.

The remaining discrepancies for L2 and L5 mark the limit of our approach. The closure in Eq.~\eqref{eq:three-point} effectively determines the third moment of the goodness distribution from the first two. Its error generally grows with the covariance $C$ but remains small when the goodness is concentrated around a single dominant value. We expect this not to be the case for L2 and L5, whose goodness distributions lack such a dominant peak, particularly at low assessment error~\cite{FujimotoOhtsuki2024}, so that information beyond the first two moments becomes relevant. A quantitative account of these higher-order correlations is left for future work. 

We focused on the assessment structure of a given population rather than on evolutionary stability against mutants, which additionally requires the payoffs and assessment dynamics of invading strategies~\cite{FujimotoOhtsuki2023,FujimotoOhtsuki2024}. Since the stability conditions of Murase and Hilbe are expressed in terms of $h$ and $h_{\mathrm G}$~\cite{MuraseHilbe2024}, the closed equations derived here could provide the input for such an analysis. Related evolutionary analyses have examined competition between behavioral strategies under fixed assessment rules~\cite{OhtsukiIwasa2007} and between different assessment rules~\cite{UchidaSigmund2010}, and recent work has studied the convergence of assessment dynamics under public and private assessment~\cite{MorskyPlotkinAkcay2024}.

Our framework can also be extended to settings that change how strongly observers agree, such as gossip and opinion synchronization~\cite{KawakatsuKessingerPlotkin2024,MuraseHilbe2024} and public reputation institutions~\cite{Radzvilavicius2021}, which bring assessments closer to the public limit, partial observation~\cite{Hilbe2018}, which reduces the number of commonly observed actions, and continuous assessments~\cite{LeeMuraseBaek2021} or alternative update schemes~\cite{MuraseHilbe2023,OkadaDeSilva2024}. In each case, the question is how the correlation between observers, and hence the level of cooperation, changes. 
\section{Methods}
\label{sec:methods}

\subsection{Simulation protocol}
\label{sec:sim-protocol}
Here, we describe the overall simulation procedure. For each realization, all individuals use the same action and assessment rules, which remain fixed throughout the simulation. At each interaction, an ordered donor--recipient pair $(d,r)$ with $d\neq r$ is sampled at random. The donor acts according to the action rule described in Section~\ref{sec:private-image-matrix}, after which all observers, including the recipient, update their assessments of the donor simultaneously using the pre-update image matrix. With probability $\epsilon$, independently for each observer, an assessment error occurs, and the observer assigns the opposite assessment to that prescribed by the norm. One Monte Carlo (MC) step is defined as $N$ donor--recipient interactions. Since the donor and recipient are sampled uniformly, each individual is selected once as donor and once as recipient on average per MC step. The observables $h$ and $h_{\mathrm G}$ are recorded at $t_{\mathrm{MC}}=0$ and after every MC step thereafter.

Unless otherwise stated, simulations use a population size of $N=\PopulationSize$ and are performed separately for each of the Leading Eight norms. Initial off-diagonal entries of the image matrix are independently set to $\Bad$ with probability $\InitialBadFraction$ (and to $\Good$ otherwise), while all diagonal entries are initialized as $\Good$ and maintained throughout the simulation. The assessment error probability $\epsilon$ is varied from $\AssessmentErrorMin$ to $\AssessmentErrorMax$ in increments of $\AssessmentErrorStep$. Each realization consists of $2N(N-1)$ donor--recipient interactions, corresponding to $2(N-1)$ MC steps. For $N=\PopulationSize$, the system is therefore simulated from $t_{\mathrm{MC}}=0$ to $t_{\mathrm{MC}}=\FinalMCTime$. We verified that $h$ and $h_{\mathrm G}$ reach stationary values well before $t_{\mathrm{MC}}=\FinalMCTime$ for all parameter values considered (see SI). For each realization, $h$ and $h_{\mathrm G}$ are computed from the image matrix at the final time, $t_{\mathrm{MC}}=\FinalMCTime$, using Eqs.~(\ref{eq:h-def}) and~(\ref{eq:hG-matrix}). The simulation values shown in Figs.~\ref{fig:steady_h} and~\ref{fig:hG-comparison} are ensemble averages of these quantities over the \NumberRuns\ realizations.

\subsection{One- and two-variable mean-field (MF) approximations}

To describe the assessment dynamics, we consider how $h$ and $h_{\mathrm G}$ change after a single donor-recipient interaction. Since one Monte Carlo step consists of $N$ interactions, each interaction advances time by $dt=1/N$, and we express $h(t+dt)$ and $h_{\mathrm G}(t+dt)$ in terms of $h(t)$ and $h_{\mathrm G}(t)$. The stationary values $\tilde h$ and $\tilde h_{\mathrm G}$ are then obtained in the limit $t\rightarrow\infty$, where $h(t+dt)=h(t)$ and $h_{\mathrm G}(t+dt)=h_{\mathrm G}(t)$.

We first consider the dynamics of $h$. After a single interaction, every observer updates its assessment of the donor, so that one column of the image matrix, i.e., a fraction $1/N$ of the off-diagonal assessments, is updated. Denoting by $F$ the probability that an updated assessment is good, we have
\begin{equation}
 h(t+dt)=\left(1-\frac{1}{N}\right)h(t)+\frac{1}{N}\,F.
 \label{eq:h-update}
\end{equation}
An observer updates its assessment of the donor depending on how it currently assesses the donor, $x_{od}$, and the recipient, $x_{or}$, as well as on the donor's action $A=a(1, x_{dr})$. Thus, in the absence of assessment errors,
\begin{equation}
 F=
 \sum_{x_{od}, x_{or},x_{dr}\in\{0,1\}}
 \Pp(x_{od}, x_{or},x_{dr})\, \nu_{A}(x_{od}, x_{or}),
 \label{eq:MF-h}
\end{equation}
where $\Pp(x_{od}, x_{or},x_{dr})$ is the joint probability of the three assessments at time $t$, and $\nu_{A}(x_{od},x_{or})$ is the assessment rule given in Table~\ref{tab:leading-eight}. Here we neglect the case in which the observer is the recipient itself, where $x_{or}$ is a diagonal element; this concerns only one of the $N-1$ updated assessments and is negligible for large $N$.
With an assessment error occurring with probability $\epsilon$, $\nu_{A}(x_{od}, x_{or})$ in Eq.~\eqref{eq:MF-h} is replaced by
\begin{equation}
 \mu_A(x_{od}, x_{or})
 =
 (1-\epsilon)\nu_A(x_{od}, x_{or})
 +
 \epsilon\left[1-\nu_A(x_{od}, x_{or})\right].
\label{eq:assessment-error}
\end{equation}

In the one-variable mean-field approximation, every off-diagonal assessment is independently $\Good$ with probability $h$ and $\Bad$ with probability $1-h$. The probability that a player assesses another player as $x\in\{0,1\}$ is then $\pi_h(x)=h^x(1-h)^{1-x}$, and the joint probability factorizes, yielding
\begin{equation}
 F(h)=
 \sum_{x_{od}, x_{or},x_{dr}\in\{0,1\}}
 \pi_h(x_{od})\,\pi_h(x_{or})\,\pi_h(x_{dr})\, \mu_{A}(x_{od}, x_{or}),
 \label{eq:MF-h-closed}
\end{equation}
which closes Eq.~\eqref{eq:h-update} in terms of $h(t)$. Taking $t \rightarrow \infty$, Eq.~\eqref{eq:h-update} reduces to the self-consistency equation $\tilde h = F(\tilde h)$, whose solution gives the stationary value.

Note that, for the L7 norm without assessment errors, $F(h)=h$, so that the stationary condition $\tilde h = F(\tilde h)$ is satisfied by any $\tilde h$: the dynamics is neutral, and $\tilde h$ is left undetermined. This indeterminacy arises from the infinite-population limit implicit in Eq.~\eqref{eq:MF-h-closed}. In a finite population of size $N$, where players are sampled without replacement, we instead obtain $\frac{dh}{dt}=\frac{h(1-h)}{N-2}$ so that $h$ increases toward $\tilde h=1$. The detailed derivation, including the case with assessment errors, is provided in the Supplementary Information.

The approximation in Eq.~\eqref{eq:MF-h-closed} is valid when the assessments $x_{od}$, $x_{or}$, and $x_{dr}$ are mutually independent. However, assessments of the same player by different observers are generally correlated, because they are updated upon observing the same action. Hence $x_{or}$ and $x_{dr}$, which both concern the recipient, would not be independent. In contrast, $x_{od}$ concerns a different player, and we assume that it remains independent of them, so that $\Pp(x_{od}, x_{or}, x_{dr})=\pi_h(x_{od})\,\Pp(x_{or}, x_{dr})$. To handle the joint probability $\Pp(x_{or}, x_{dr})$, we introduce the conditional probability $h_{\mathrm G}$ ($h_{\mathrm B}$) that an observer regards a player as $\Good$ given that another observer regards the same player as $\Good$ ($\Bad$). We then write
\begin{equation}
\Pp(x_{or}, x_{dr}) = \pi_h(x_{dr})\,\pi(x_{or}\,|\,x_{dr}),
\label{eq:two-point}
\end{equation}
where $\pi(x\,|\,1)=\pi_{h_{\mathrm G}}(x)$ and $\pi(x\,|\,0)=\pi_{h_{\mathrm B}}(x)$. The probability that the first of two observers regards a player as $\Good$ and the second as $\Bad$ equals that of the opposite case, $h(1-h_{\mathrm G})=(1-h)h_{\mathrm B}$, so that $h_{\mathrm B}$ is not an independent variable but is expressed in terms of $h$ and $h_{\mathrm G}$:
\begin{equation}
 h_{\mathrm B}
 =
 \frac{h(1-h_{\mathrm G})}{1-h}.
 \label{eq:hB}
\end{equation}
Using Eq.~\eqref{eq:two-point} in Eq.~\eqref{eq:MF-h}, we obtain the two-variable mean-field expression
\begin{equation}
 F(h, h_{\mathrm G})=
 \sum_{x_{od}, x_{or},x_{dr}\in\{0,1\}}
 \pi_h(x_{od})\,\pi_h(x_{dr})\,\pi(x_{or}\,|\,x_{dr})\, \mu_{A}(x_{od}, x_{or}),
 \label{eq:MF-g-two}
\end{equation}
whose right-hand side depends on both $h(t)$ and $h_{\mathrm G}(t)$. To close Eq.~\eqref{eq:h-update}, we therefore need the dynamics of $h_{\mathrm G}$.

The dynamics of $h_{\mathrm G}$ is obtained in a similar way. The difference is that we now consider two distinct observers, labeled 1 and 2, and the joint probability $h\,h_{\mathrm G}$ that both regard the same player as $\Good$. Since only the assessments of the donor are updated,
\begin{equation}
 h(t+dt)\,h_{\mathrm G}(t+dt)=\left(1-\frac{1}{N}\right)h(t)\,h_{\mathrm G}(t)+\frac{1}{N}\,Q,
 \label{eq:hG-update}
\end{equation}
where $Q$ is the probability that both observers regard the donor as $\Good$ after the update. This probability depends on five assessments, $x_{dr}$, $x_{1d}$, $x_{1r}$, $x_{2d}$, and $x_{2r}$, where an integer subscript labels an observer; e.g., $x_{1r}$ is observer 1's assessment of the recipient. Then,
\begin{equation}
 Q=
 \sum_{\substack{x_{dr},x_{1d}, x_{1r},\\x_{2d}, x_{2r}\in\{0,1\}}}
 \Pp(x_{dr},x_{1d}, x_{1r}, x_{2d}, x_{2r})\, \mu_{A}(x_{1d}, x_{1r})\,\mu_{A}(x_{2d}, x_{2r}),
 \label{eq:MF-hG}
\end{equation}
where $\Pp(x_{dr},x_{1d}, x_{1r}, x_{2d}, x_{2r})$ is the joint probability of the five assessments at time $t$. As before, we neglect the cases in which observer 1 or 2 is the recipient, which are negligible for large $N$.

Assuming again that assessments of different players are independent, the joint probability factorizes as $\Pp(x_{dr},x_{1r}, x_{2r})\,\Pp(x_{1d}, x_{2d})$. The second factor is a two-point probability and is given, as in Eq.~\eqref{eq:two-point}, by $\pi_h(x_{1d})\,\pi(x_{2d}\,|\,x_{1d})$. The first factor, however, involves a three-point correlation, and describing it would in turn require still higher-order correlations. To close the equations at the level of $h$ and $h_{\mathrm G}$, we approximate
\begin{equation}
 \Pp(x_{dr},x_{1r}, x_{2r})\approx \pi_h(x_{dr})\,\pi(x_{1r}\,|\,x_{dr})\,\pi(x_{2r}\,|\,x_{dr}),
 \label{eq:method-three-point}
\end{equation}
that is, we assume $\Pp(x_{2r}\,|\,x_{1r},x_{dr})\approx \Pp(x_{2r}\,|\,x_{dr})$. We thus obtain
\begin{equation}
\begin{aligned}
 Q(h, h_{\mathrm G})=
 \sum_{\substack{x_{dr},x_{1d}, x_{1r},\\x_{2d}, x_{2r}\in\{0,1\}}}
 &\pi_h(x_{1d})\,\pi(x_{2d}\,|\,x_{1d})\,
 \pi_h(x_{dr})\,\pi(x_{1r}\,|\,x_{dr})\,\pi(x_{2r}\,|\,x_{dr})\\
 &\times \mu_{A}(x_{1d}, x_{1r})\,\mu_{A}(x_{2d}, x_{2r}).
 \end{aligned}
 \label{eq:MF-hG-closed}
\end{equation}
Taking $t\rightarrow\infty$ in Eqs.~\eqref{eq:h-update} and \eqref{eq:hG-update} with Eqs.~\eqref{eq:MF-g-two} and \eqref{eq:MF-hG-closed}, the stationary values are obtained by solving
\begin{equation}
 \tilde h = F(\tilde h, \tilde h_{\mathrm G}),
 \qquad
 \tilde h\, \tilde h_{\mathrm G} = Q(\tilde h, \tilde h_{\mathrm G}).
 \label{eq:stationary}
\end{equation}

\subsection{Code availability}
The simulation, theoretical calculation, and figure-generation codes used in this study are publicly available at \url{https://github.com/DynamicCube-lab/Indirect_reciprocity}.

\section*{Data availability}
The datasets generated and analysed during this study, including simulation outputs and numerical results supporting the figures, are publicly available at \url{https://github.com/DynamicCube-lab/Indirect_reciprocity}.

\bibliography{bibliography}

\backmatter

\section*{Acknowledgements}

This research was supported by the National Research Foundation of Korea (NRF) grant Nos. RS-2024-00460958 and RS-2026-25497103 (H.J.P.)

\section*{Author contributions}

J.-W.H. performed the numerical analysis and MF calculations and wrote the manuscript. S. C. supervised the study and revised the manuscript. H.J.P. designed the project and supervised the study and revised the manuscript. 

\section*{Competing interests}

The authors declare no competing interests.

\end{document}